\documentclass[aps,pre,twocolumn,showpacs,floatfix]{revtex4}

\usepackage{amsmath}
\usepackage{graphicx}

\newcommand{\etal}{{\it et al. }}
\newcommand{\ie}{{\it i.e.}}
\newcommand{\vfc}{\ensuremath{\phi}}
\newcommand{\vfp}{\ensuremath{\phi_p}}

\newcommand{\func}[2]{#1\!\left(#2\right)}

\renewcommand{\vec}[1]{\ensuremath{\boldsymbol{#1}}}
\newcommand{\taur}{\ensuremath{\tau_R}}
\begin{document}

%Title of paper
\title{Brownian Dynamics Simulations of Aging Colloidal Gels}

\author{Rodolphe J.M. d'Arjuzon}
\email[]{rjmd2@phy.cam.ac.uk}

\affiliation{Cavendish Laboratory, University of Cambridge,
Madingley Road, Cambridge CB3 0HE, U.K.}

\author{William Frith and John R. Melrose}
\affiliation{Colworth House, Unilever Research, Bedford, U.K.}

\date{\today}

\begin{abstract}    
Colloidal gel aging is investigated using very long duration
brownian dynamics simulations. The Asakura Oosawa description of the depletion
interaction is used to model a simple colloid polymer mixture. Several regimes
are identified during gel formation. The Intermediate scattering function
displays a double decay characteristic of systems where some kinetic
processes are frozen. The $\beta$ relaxation at short times is explained in
terms of the Krall-Weitz model for the decorelation due to the elastic modes
present. The $\alpha$ relaxation at long times is well described by a stretched
exponential, showing a wide spectrum of relaxation times for which the $q$
dependence is $\tau_{\alpha} = q^{-2.2}$, lower than for diffusion. For the
shortest waiting times, a combination of two stretched exponentials is used,
suggesting a bimodal distribution. The extracted relaxation times vary with
waiting time as $\tau_{\alpha}=\tau_w^{0.66}$, slower than the simple aging
case. The real space displacements are found to be strongly non-Gaussian,
correlated in space and time. We were unable to find clear evidence that the gel
aging was driven by internal stresses. Rather, we hypothesise that in
this case of weakly interacting gels, the aging behaviour arises due to the
thermal diffusion of strands, constrained by the percolating network which
ruptures discontinuously. Although the mechanisms differ, the similarity of
some of the results with the aging of glasses is striking.  \end{abstract}

% insert suggested PACS numbers in braces on next line

\maketitle

\section{introduction}

Like other non-equilibrium phenomena, colloidal gelation has attracted
considerable interest in the last twenty years. Colloids are found in many
forms and are highly relevant for industry. Muds, paints, cosmetics, and many
food products are examples where understanding of colloid science is essential
in the design of product properties. Studies can be conducted over a wide range
of densities, and the interactions can be finely tuned in terms of range and
depth of potential by altering the particle coating, solvent conditions, or
solutes present.

Importantly, under certain conditions it is possible to make colloid particles
form a macroscopic gel. Although there is no universal definition, a gel is
usually understood to be a percolating viscoelastic  network that has no long
range order (although it may have some local order), but a characteristic
length scale is present; gels are usually fractals at low volume fractions but
this is not a necessity. Common examples of colloid gels include yogurt,
toothpaste, and clays.

The system under study here is a colloid polymer mixture. Such systems have
been known to have interesting equilibrium and non-equilibrium properties for
many years \cite{poon95}; they are also a stepping stone to understanding the
more complicated mixtures that make up many ordinary products: shampoo and
paint are examples which involve complex polymer colloid mixtures. Such systems
can display depletion flocculation, and the resulting gels are often used as
one of the simplest models of weakly attractive system due to their relative
simplicity arising from the absence of charge effects.

The properties of non-equilibrium systems as they age are not only relevant for
applications, but are also a crucial step to understanding the very nature of
gels. Indeed if slow aging were not a feature of the gels, phase separations
would only result in equilibrium phases.

The dramatic slow down of the dynamics in gels  has prompted investigations
into their possible link with glasses. While gels are usually found at lower
volume fractions and when strong interparticle  interactions are present,
colloid glasses are usually found at high volume fractions, above \vfc = 0.58,
or for deep quenches. They share many properties such as being out of
equilibrium, non-ergodic, the absence of long range order and slow dynamics.
Since work on the glasses and their aging is extremely topical in soft
condensed matter, this work will look for  similarities between the aging of
colloid gels and glasses.

There is abundant literature on gels and on colloid polymer systems in
particular. Asakura and Oosawa were the first to describe the attraction between
colloid particles when non-adsorbing polymer is present in solution \cite{ao}.
The depletion mechanism which is responsible for this can be understood in
different ways. The essential point is the presence of a region around the
colloids from where the polymer coil centre of mass is excluded. When two
colloid particles come together the regions of excluded volume overlap
resulting in an increase of the free volume available to the polymer, causing
an increase of their entropy and an overall decrease of the free energy of the
system. The problem can also be analysed in terms of the net osmotic pressure
on a pair of colloids as they come together.

Theoretical work on the system was done by Vrij \cite{vrij}, who independently
rediscovered the Asakura and Oosawa model, and demonstrated the presence of
spinodal instability in the system. Gast \etal treated the polymer as a
perturbation to the hard sphere results thus producing the first phase diagrams
\cite{gast83,gast85}. Lekkerkerker \etal \cite{lek92} used a free volume
approach to treat this problem while Djikstra \etal formally integrated out
polymer degrees of freedom \cite{dj}. It has been shown that for size ratios
below $k=0.154$ there can be no three body interactions and that the two body
Asakura and Oosawa (AO) potential can therefore be an adequate description. A
review of the details of the different approaches to the colloid polymer system
can be found in \cite{dj}. It is important to note
that the depletion potential give well controlled systems where the
interparticle interaction potential well is usually in the range of $1-10kT$,
very different from emultions, for example, where interaction energy can be of
order $100kT$.

Felicity \etal \cite{heyes95} and Lodge \etal \cite{heyes99} modelled colloid 
gels interacting via a Leonard-Jones potential and studied their viscoelastic
properties \cite{heyes99,heyes99b}; Bijsterbosh \etal investigated fractal
colloidal gels for various interactions using brownian dynamics simulation
methods \cite{dick95}. Dickenson \etal \cite{dick00}, and Bos \etal
\cite{bos96} have worked on stress relaxations after shear stresses were
applied. Soga \etal have worked on percolation \cite{sogaperc} and phase
transitions \cite{sogaphase} in the colloid polymer system. The yielding
behaviour of 2D colloid gels was investigated by West \etal \cite{west94}. 

Experimentally, most notable was the work of Grant and Russel \cite{grant93},
Poon \etal \cite{poon95}, Lekkerkerker \etal \cite{lek95}, and Verdin and
Dhont\cite{dhont95}. An important result from all these works was that at
moderate volume fractions $(0.1<\vfc<0.4)$ gelation can be expected to occur
for high polymer concentrations (deep quenches). Also, for ranges of
interaction of less than a third of the particle diameter no liquid phase
formed; the liquid gas binodal becoming metastable with respect to fluid solid
phase separation. In light scattering, the gelation is marked by the `freezing'
of the low q peaks demonstrating that the dynamics become arrested as the sol
gel transition is reached.

The following experiments are explained in more detail since they contain the
most interesting data for comparison with our simulations. Cipelletti \etal
\cite{cipe1} have produced experimental results on the aging of polystyrene
colloid gels at low volume fractions ($10^{-4}<\phi<10^{-3}$). The gels where
formed by aggregation due to the van der Walls forces, and were found to be
fractal. The aging behaviour was sampled for waiting time going from hours to
ten days, the relaxations where stretched with the exponent greater than one.
The unusual $q$ dependence of the relaxation times, $\tau_f \propto q^{-1}$,
was explained in terms of a dipole elastic field caused by the high internal
stresses of the gels. The waiting time dependence of the relaxation times was
nearly linear, $\tau_f \propto \tau_w^{0.9}$, following a period of exponential
growth (this exponential regime was also observed in an aging clay suspension
\cite{abou01}). 

The same group, in a series of rheology and dynamic light scattering
experiments, also found the same behaviour in a system composed of
multilamellar vesicles, where, once again, it was thought to be the strong
internal \emph{repulsive} stresses that dominate the response \cite{cipe2}. 

Experiments on the dynamics of colloidal gels have been performed by Krall and
Weitz \cite{krall98} for the short time (or $\beta$, see below) relations. They
also presented a model for the elastic modes present in a fractal gel, from
which the mean squared displacements of strands constrained by the network can
be calculated. Working on gel at densies conparable with this work, Romer \etal
\cite{romer00}, showed that the predictions of this model held even for
non-fractal structures. 

Both short and long time scale dynamics where observed by Solomon \etal
\cite{solomon01} in their study of adhesive colloids at low volume fractions
and for interparticle interactions strengths close to that used in this work
(~$10kT$). The Krall-Weitz model was used at short times, which a simple
exponential was observed at long times. 

A commonly calculated quantity is the intermediate scattering function (ISD), a
time dependent correlation function. When looking at glasses and gels the self
part of the ISD will display a double decay. At short times a fast decay often
referred to as the $\beta$ relaxation, and at long times a second decay called
the $\alpha$ relaxation. Structural arrest, brought about by increased density
for hard spheres or by quenching for a Leonard-Jones system, will lead to a
wide separation of these time-scales and the appearance of a plateau at
intermediate times see reference \cite{kob} and references therein. Since the
scattering function is wavevector dependent, spatial information can also be
obtained. 

The $\alpha$ and $\beta$ relaxations are often explained in terms of the cage
effect. For a system of purely repulsive particles, a particle is confined by
its neighbours which form a cage around it; the particle is free to diffuse
within its cage, thus setting the $\beta$ time scale. The time required for the
particle to diffuse out of the cage is the $\alpha$ time scale. As a system
becomes denser, cage rearrangements become increasingly difficult, causing the
$\alpha$ relaxations to occur on longer time scales. For systems of attractive
particles, one can talk about an open cage, whereby a particle would be
confined to its neighbours because of the attractive forces; two relaxation
time scales also result from this. On this basis it has been argued that the gel
transition 
should be described in terms of Mode Coupling theory (MCT) as an ergodicity
breaking transition
\cite{berg1,berg2,berg3,puertas02}. MCT has considerably helped to
understand supercooled liquids \cite{kob}, and high volume fraction colloid
glasses \cite{pusey86}, and its predictions are now being tested for gels as
well \cite{solomon01}. The kinetic arrest of gels has also been reported by
Segr\`e \etal, who compared it to the features of the hard sphere colloid glass
transition \cite{segre01}. 

Ideal MCT assumes closeness to equilibrium and is therefore only useful for
the fluid-gel transition. Non-equilibrium MCT has also been developed by
Bouchaud \etal \cite{bouchaud96} which predicts aging behaviour. Some of these
predictions have been qualitatively tested on a Laponite glass \cite{abou01}.
However, earlier theoritical models have interesting aging properties.
Cugliandolo and Kurchan have solved the spin glass model for the out of
equilibrum dynamics \cite{cugliandolo93}. Crucially, these equations are
identical to those produced by some MCT models.

Another approach to explain the behaviour of glasses has been the study of the
properties of the Potential Energy Surface (PES) as proposed by Goldstein
\cite{golstein69}. One model which uses this concept is the successful trap
model of Bouchaud \cite{bouchaud92} which has been widely applied for spin
glasses and of which many variants exist, this was one of first models to make
useful predictions about aging. Another formalism is now introduced to
facilitate discussion. The Inherent Structures (IS) of a system are defined as
the local minima of the PES; all points in configuration space which end up in
the same IS  following a steepest decent path define the basin of that IS
\cite{stillinger95}. Sciotino and Tartaglia distinguish between four dynamical
regimes: a high temperature region where the system is fluid; an intermediate
region where the system is always close to the ridges separating the different
basins, this being the regime in which MCT is successful; the third regime is
found at low $T$ where the system populates IS with lower and lower energy.
When the time of escape from a basin becomes the longer than the experimental
time scale the system no longer equilibriates, which is the fourth regime,
normally called a glass \cite{sciortino01}. In this formalism the two step
relaxation of the correlation function is seen as decoupling of the fast
(intra-well) vibrational excitations and the slower rearrangements
corresponding to the system sampling different basins which for low $T$ may
require activated processes. 

It is useful to set up a few hypothesises about the possible causes of the
aging in gels. Firstly, very slow crystallisation about some locally highly
ordered regions is a possibility although this is not normally seen in
experiments. Secondly, the aging may be stress driven as argued to be in some
experimental systems \cite{cipe1,cipe2}. Thirdly, thermal fluctuations and
random breaking of bonds to allow the relaxation of the network as suggested by
Solomon \etal \cite{solomon01}. Compactification of strands up to a point where
they might dislocate has also been suggested as a possibility to explain the
collapse of transient gels \cite{poon99sed}.

\section{Methods}
\subsection{The model}
The Brownian Dynamics model is based on a Langevin type equation of motion for
each colloid particle with a fluctuating random force accounting for the thermal
collisions of the solvent molecules with the particle.  The polymer coils are
not simulated, their presence only being felt through the attractive potential
between the colloids.  

\begin{equation}
	M \ddot{\vec{X}} = \vec{F}^{C} +
	\vec{F}^{B} + \vec{F}^{D} = 0
\end{equation}

Where $M$ and $\vec{X}$ are the inertia and position matrices respectively;
$\vec{F}^{C}$ are the conservative forces, $\vec{F}^{B}$ are the random brownian
force, and $\vec{F}^{D}$ are the dissipative drag force. This last term
encompasses all the hydrodynamic interactions, which couples the forces on the
particles through the solvent. It is these forces which oppose the relative
motion of the particles, and are most important in flow. They will affect
non-equilibrium phenomena at rest, but in common with many others we neglect
the many body interactions in this study. In this model, this term is reduced
to Stoke's drag for a spherical particle ($\alpha)$, applied individually: this
is the free draining approximation. So the drag force on each particle
satisfies:

\begin{equation}
   \vec{f_i}^{D}  =  -\alpha \vec{v_i} = -6\pi\eta \sigma \vec{v_i}
\end{equation}
where $\eta$ is the viscosity of the liquid, $\sigma$ is the particle diameter
and $\vec{v}_i$ is is the velocity of particle $i$. The inertia terms are neglected
so this becomes a force balance equation, which is solved for the displacements
of the particles. The conservative forces are the sum of a steep core
potential, and the depletion force obtained via the following two body
potential energy term.

\begin{equation}
	U = \frac{1}{2} \sum_{ij}
	\Big(
        \func{U_{HC}}{r_{ij}} + \func{U_{AO}}{r_{ij}}
	\Big),
	\label{conpot}
\end{equation}
where the first term, $U_{HC}$, is the power law repulsive hard core,

\begin{equation}
	\frac{U_{HC}}{kT}  =  \left(\frac{r_{ij}}{\sigma}\right)^{-n}.
	\label{hardcore}
\end{equation}
n = 36 is used to avoid anomalies that may result from having a softer
potential \cite{melrose92} while retaining computer time efficiency.
The two-body depletion potential used was:

\begin{equation}
	\frac{U_{AO}}{kT} = Q \func{H}{L,r_{ij}}
	\left(
	L^2\frac{r_{ij}}{\sigma} -
	\frac{1}{3}\left(\frac{r_{ij}}{\sigma}\right)^3 - A
	\right),
\end{equation}
Where $r_{ij}$ is the separation between spheres i and j. If $k$ is the size
(diameter) ratio between a polymer and a particle at the polymer volume
fraction \vfp, then
\begin{eqnarray}
	L & = & 1 + k \nonumber \\
	A & = & \frac{2L^3}{3} \nonumber \\
	Q & = & \frac{3\vfp}{2k^3} \nonumber \\
	\func{H}{L,\frac{r_{ij}}{\sigma}} & = & \left\{
	\begin{array}{ll}
		1, & \frac{r_{ij}}{\sigma} < L\\
		0, & \frac{r_{ij}}{\sigma} > L .
	\end{array}
	\right.
\end{eqnarray}
Strictly speaking, the parameter
$\vfp = \frac{1}{6}\pi k^2\sigma^2n_R$,
where $n_R$ is the polymer number density in a reservoir of pure polymer which
would match the chemical potential of the polymer solution with colloid
particles at volume fraction $\vfc$ \cite{lek92}.
It useful that the potential used here is of finite range, as this avoids
truncation errors and allows us to define nearest neighbours and interacting
particles precisely.

The Brownian forces represent the action of the liquid molecules averaged out
over one time step. They average out to zero, and are uncorrelated in time.

\begin{eqnarray}
	<\vec{F}_B> & = & 0 \\
	<\vec{F}_B(t)\vec{F}_B(t')> & = & 2\pi \alpha k_BT\delta(t-t')
\end{eqnarray}
The fluctuation-dissipation theorem relates the magnitudes of $\vec{F}_B$ and
$\alpha$ by enforcing that
every squared separable degree of freedom satisfies the equipartition theorem.

In order to simplify the calculation reduced units are used in the simulation,
$k_BT$ becoming the energy unit.
The polymer volume fraction, $\phi_p$, is the only control parameter for the
depth of potential, and  the ratio of the colloid to polymer size $k$
controls the range of the potential. This was fixed at $k = 0.1$ for all
simulations shown here.

All distances are measured in units of the particle diameter $\sigma$.

The natural time unit to use is the Brownian relaxation time. It is defined as
the time for a particle to diffuse its own length scale by Brownian motion. 

\begin{equation}
\tau_R = \frac{\eta \sigma^3}{k_BT}
\end{equation}
For a micron size particle in water, \taur\ is of order 0.2 seconds. Note
however the strong dependence on particle size. The typical length of runs
showed here was of 5000 \taur\ and the longest run was twice that. This is a
great increase compared to previous studies which were typically run for about
100 \taur.

The system being simulated is a three dimensional cubic box with 4000 particles
in Periodic Boundary Conditions.

\subsection{Quantities of interest}

As mentioned earlier the self part of the intermediate scattering function
(also called incoherent intermediate scattering function) gives information
about the relaxation of density fluctuations in the sample. It lets us assess
whether particles are displaced on a length scale set by $\frac{2\pi}{\vec{q}}$
or whether kinetic processes of that wavelength are 'frozen out', it is
a commonly calculated quantity as it can be measured in light scattering
experiments \cite{cipe1}. It is calculated from simulation data by 

\begin{equation}
F_s(q,t,t_w) = \frac{1}{N} \sum_{i} e^{i \vec{q} \cdot \left(
		\vec{r_{i}}(t_{w}) - \vec{r_{i}}(t_w + t) \right)},\label{intscaeq}
\end{equation}
where $t_w$ is the waiting time for that sample.
The wave vector $\vec{q}$ is discretised to correctly take into account the
periodic boundary conditions. Therefore, for each axis we have:

$$ q  =  \frac{2\pi n}{L}, n  =  0, 1, 2, ...$$   
with $L$ the linear dimension of the box. To improve the statistics the results
are averaged over the x, y and z axes. Since it depends both on $t$ and $t_w$,
$F_S$ is a so-called {\it two time quantity}, which are most useful in
determining and characterising any aging behaviour if it is present.

In order to calculate particle correlations in real space the radial
distribution function g(r) is calculated, as well as the self part of the Van Hove
correlation function which samples particle displacements in time. 

$$ g(r) = \frac{1}{<n>}\sum_{i\ne j}^N \delta(|\vec{r}_i-\vec{r}_j| - \vec{r}) $$
where $<n>$ is the average number density,

$$ G_s(r,t) = \frac{1}{N}\sum_i^N < \delta(|r_i(t)-r_i(0)| - r) > $$ 
where $<...>$ is an average over time origins.
The quantity

$$ S(r,t) = 4\pi r^2 G_s(r,t) $$ gives the probability of finding a particle at
a distance r after a time t, and $G_s(r,0)=g(r)$.

In the Gaussian approximation, $$G_s(r,t) = \frac{1}{(4\pi Dt)^{\frac{3}{2}}}\ exp
\left( \frac{-r^2}{4Dt} \right) $$ which is a solution of the diffusion equation.

Our results are divided into two parts; the first part will be concerned with
`macroscopic' quantities, \ie\ correlation functions and quantities that have
been averaged over several systems; the second part will look more
closely at the `microscopics' of individual systems.

\section{Results}
\subsection{Macroscopics}

As in previous simulations \cite{sogaperc,sogaphase} the samples were
equilibrated at high temperature (\vfc = 0) and instantaneously quenched to an
appropriate value of \vfp. Work in \cite{sogaphase} determined that for \vfp
$\geq 0.3$ crystals were no longer obtained. For these relatively deep
quenches, amorphous, non-equilibrium structures were obtained; these networks
come under the general name of gels or transient gels when seen experimentally
before they collapse gravitationally. They are qualitatively similar to created
experimentally gels and LJ gels from simulations. All gels considered here have
\vfc\ = 0.30 and \vfp\ = 0.40, giving a potential minimum of $U_{min}=-8kT$.
Properties are averaged over five samples unless otherwise stated. 

  \begin{figure}[t]\centering\footnotesize
  \includegraphics[width=6cm,angle=-90]{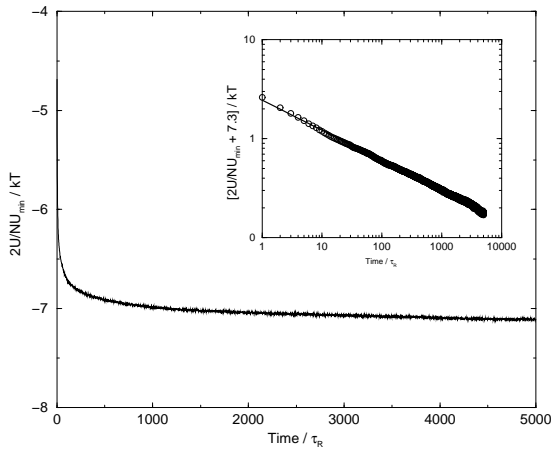}
  \caption{Potential Energy of gel, from formation to aging. The fit for the inset
  is a power law.\label{energy}} 
  \end{figure}

In figure \ref{energy} the potential energy of the system, $U$, is scaled by $U
\rightarrow \frac{2U}{NU_{min}}$ giving the approximate number of neighbours
assuming inter-particle distances are near the potential minimum. The  gel
particles have approximately seven neighbours. It is immediately obvious that
we are dealing with a non-equilibrium system since, after the initial very fast
drop in energy after quenching, a long regime of slow decay is entered; no
equilibrium value was ever reached in the simulations we ran. Such small decays
can be fitted with a number of functions, however for comparison with Barrat
and Kob \cite{barrat97}, it was fitted with a power law $U = a\cdot t^{-b}$, as
shown in the inset of figure \ref{energy}, yielding $b=0.3$.

  \begin{figure}[t]\centering\footnotesize
  \includegraphics[width=6cm,angle=-90]{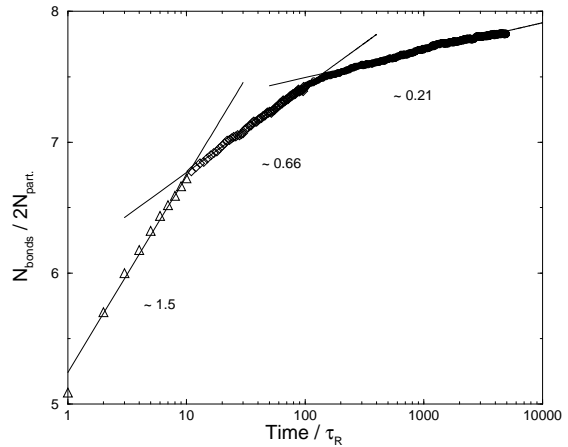}
  \caption{Total number of bonds in the gel. The three straight lines and their
  gradient are superposed to highlight the three different logarithmic regimes.
  \label{totbonds}} 
  \end{figure}
  
It is revealing to look at the total number of bonds in the system as gelation
occurs. When plotted on log-linear scale (figure \ref{totbonds}) three different
growth regimes are evident, the first up to $10 \tau_R\ $, another from $10
\tau_R\ $ to around $100 \tau_R\ $, and the last one from there on. Note that
since the ordinate range is small, a log-log plot shows practically the same
result, hence one cannot easily distinguish between a low power law growth and
logarithmic growth. The fits are of the form $N = c.log(t) + d$ where the
values of $c$ are shown on the plot. As explained in the discussion we are
here mostly concerned with behaviour over the last of these regimes.

\begin{figure}[t]\centering\footnotesize
 \includegraphics[width=6cm,angle=-90]{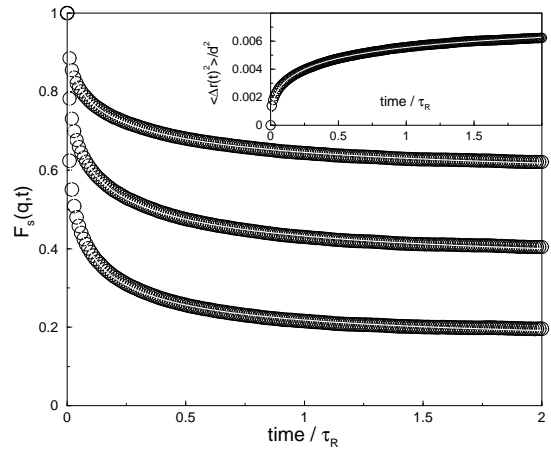}  
 \caption{Beta
 relaxation at short times for $qd = 23.0, 32.9, 46.0$, with stretched
 exponential fits given in text (white line). Inset: Mean Squared Displacement
 (black circles) for the same time scale with fit to equation \ref{deltar}
 (white line).\label{betafits}}  
 \end{figure}

The ISF was found to display the expected two step relaxation at short ($\beta$
relaxation) and long ($\alpha$ relaxation) times. They are treated separately
here since the averaging cannot be done in the same way. For the waiting time
independent $\beta$ relaxation an average over time origins can be used,
whereas for the $\alpha$ relaxation the non-ergodic nature of the system means
only an ensemble average can be performed. Figure \ref{betafits} shows this
initial decay for increasing values of $q$. As the gel forms, the mean squared
displacements of the particles is Fickean, but as aggregation proceeds the
behaviour becomes subdiffusive ($ <\Delta r(\tau)^2>\ \propto t^e,\ e < 1$).
Finally, for measurements in the region of interest the mean squared
displacement takes on the form of figure \ref{betafits} (inset) where a plateau
is reached at displacements that are small compared to the particle size. This
is interpreted in terms of the Krall-Weitz model for the initial decay of the
correlation function \cite{krall98}. This model considers the relaxation due to
the elastic modes present in the gel, with the particles assumed to be confined
to the gel strands. Their conclusion was that the mean squared displacement
should obey: 

\begin{equation}
<\Delta r(\tau)^2> = \delta^2[1-e^{(-\frac{\tau}{\tau_{\beta}})^p}] \label{deltar}
\end{equation}

The particle displacements are constrained to an amplitude $\delta$; the
stretched exponential form comes about from the sum of the relaxations on all
length scales. 

  \begin{figure}[t]\centering\footnotesize
      \includegraphics[width=6cm,angle=-90]{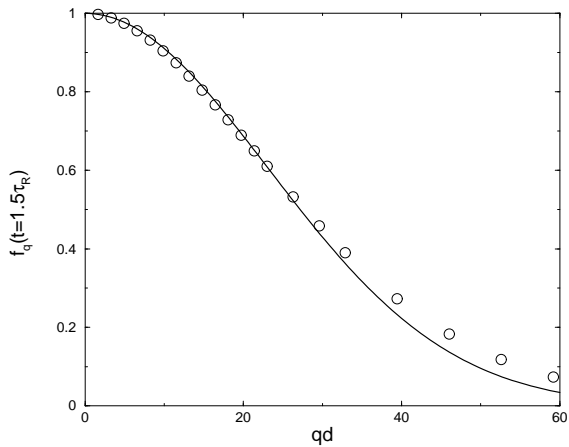}
  \caption{Non-ergodicity parameter as a function of q, with exponential fit.
  \label{fq}}
  \end{figure}

Consistently with the model we are using, the $\beta$ decay was fitted with the
following law which is closely related to a stretched exponential, 

\begin{equation}
F_s(\vec{q},t) = \exp(q^2<\Delta r^2(t)>/6) \label{stretched}
\end{equation} 

for all \vec{q} vectors, with $<\Delta r^2(t)>$ from equation \ref{deltar}. The
results are shown in figure \ref{betafits} for three different values of
\vec{q}. The value of the exponent was determined from the fits to data in
figure \ref{betafits} with equations \ref{deltar},\ref{stretched} to be $p=0.4 \pm
0.1$. Similarly, $\delta = 0.075 \pm 0.004d$ and $\tau_{\beta} = 0.26 \pm
0.04\taur$. 

$\delta^2$ being the maximum mean squared displacement (excluding any $\alpha$
processes), therefore sets the height of the plateau in $F_s$ via $F_s(q,t) =
\frac{1}{N}\sum_i e^{-q^2(\Delta r_i^2)/6} = f^s_q$,  where $f^s_q$ is the
non-ergodicity parameter for the incoherent scattering. Figure \ref{fq} shows
the non-ergodicity parameter decays exponentially for increasing $q$;
$\delta^2$ can then be extracted since $f_q = e^{-(q\delta)^2/6}$ in the
gaussian approximation. The previously extracted value of $\delta$ was used for
the fit. However there is a small ambiguity about where exactly $f_q$ should be
measured, since $\alpha$ processes are also present there is no true plateau;
in this case the values of $f_q$ were measured after $1.5\tau_R$.

  \begin{figure}[t]\centering\footnotesize
  \includegraphics[width=6cm,angle=-90]{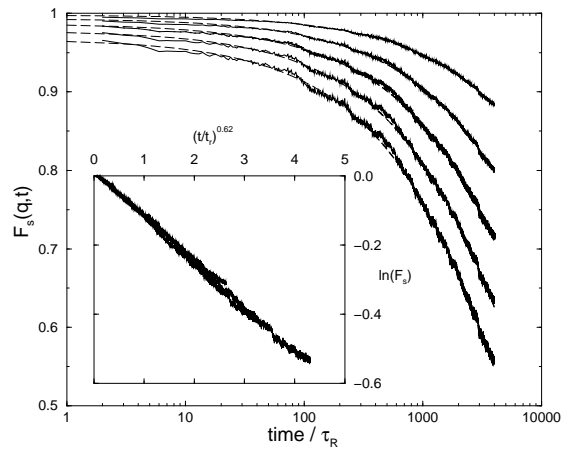}
  \caption{q vector dependence of the intermediate scattering function for long
  times and fits to
  equation \ref{stretched}. From top
  curve qd=2.30, 3.28, 4.27, 5.26, 6.25. The waiting time is 1000 \taur. Inset
  collapse of the same data when scaled as described in text and plotted against
  stretched time.
  \label{diffqcollapse}} 
  \end{figure}

We now turn to the long time dynamics; the second decay (alpha relaxation) is
the part of the dynamics which is expected to show signs of aging. Because the
system is non-ergodic, an ensemble average cannot be replaced by a time
average. Unfortunately, very long runs are needed in order to get to the time
scales that are relevant for this study, and therefore only a limited number of
runs can be performed due to computer time limitations. The averages were
therefore performed over a set of five runs. Information about the dynamics on
different wavelengths can be extracted by looking at data from a fixed waiting
time and sampling at different $q$. For a given $q$ the waiting times were then
varied to observe the aging in the dynamics.

  \begin{figure}[t]\centering\footnotesize
      \includegraphics[width=6cm,angle=-90]{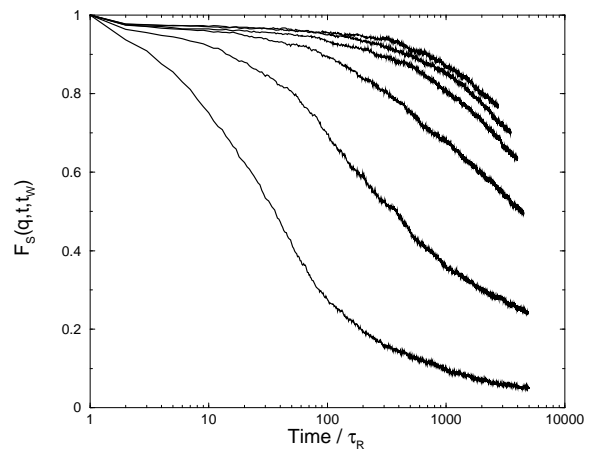}
  \caption{Intermediate scattering function at fixed qd = 6.25, and increasing waiting
  times, from left $t_w$ = 20, 100, 500, 1500, 2250 \taur.\label{avintscaaging}}
  \end{figure}

Figure \ref{diffqcollapse} shows the alpha relaxations for a single waiting
time and the $q$ vectors $qd=2.3, 3.28, 4.27, 5.26$, and $6.25$. The
relaxations were fitted to a stretched exponential law:

\begin{equation}
F_s(q,t) = A \exp(-((t-t_w)/\tau_{\alpha})^{\mu}) \label{strexp}
\end{equation}

where $A$ where was a constant chosen to be the plateau height and $\mu$ was
kept constant for different $q$. The data collapses onto a master curve
when plotted as a function of stretched time $(t-t_w)^{\mu}$ once normalised by 
$A$ and $\tau_q$ (inset of figure \ref{diffqcollapse}). The relaxation times
$\tau_{alpha}(q)$ which are extracted from the fits for different $q$ where found to
follow a power law relationship $$\tau_{\alpha} \sim q^{-\nu}$$ with $\nu=\nu(t_w)$.
The best fit values extracted were slightly age dependent as
summarised in Table \ref{tablexpo}.

\begin{LARGE}
\begin{table}[h]
  \begin{center}
  \begin{tabular}{|c|c|c|c|c|c|c|}
  \hline
  $t_w$/\taur & $\mu$ & $\nu$ & Double Exp. & $\mu_1'$ & $\mu_2'$ & $\nu_2'$ \\
  \hline \hline
    100 & - & - & Y & 0.7 & 0.7 & 2.2\\
    \hline 
    200 & - & - & Y & 0.85 & 0.7 & 2.1\\
    \hline 
    500 & 0.45  & 3.0 & Y/N & 0.9 & 0.7 & 2.2\\
    \hline 
    750 & 0.50  & 2.7 & N & - & - & -\\
    \hline 
    1000& 0.62  & 2.5 & N & - & - & -\\
    \hline 
    1250& 0.70  & 2.3 & N & - & - & -\\
    \hline 
    1500& 0.70  & 2.3 & N & - & - & -\\
    \hline 
    2000& 0.72  & 2.2 & N & - & - & -\\
    \hline 
  \end{tabular}
  \end{center}
  
  \caption{Summary of the exponents as a function of age of the gel\label{tablexpo}}
\end{table}
\end{LARGE}

For waiting times $t_w \le 1000\ \taur$, as well as fast decreasing values of
$\mu$, it was noticed that the quality of the fits decreased becoming
impossible for $t_w < 500\ \taur$. The $\alpha$ relaxation at these early
times displayed a relatively fast decay followed by a slower one (see curves
for $t_w=20,100\ \taur$ in figure \ref{avintscaaging}. These relaxations were
fitted with a double stretched exponential: 

\begin{eqnarray}
F_s(q,t) & = & \frac{f_q}{(1+A_1)}\Biggl\{\exp\left(-\left(\frac{t-t_w}{\tau_{\alpha
1}'}\right)^{\mu_1'}\right) \nonumber\\
& & +A_1 \exp\left(-\left(\frac{t-t_w}{\tau_{\alpha2}'}
\right)^{\mu_2'}\right)\Biggr\}\label{doublestrexp}
\end{eqnarray}

The justification for this law is given in the discussion section. Once again
the relaxation times contain wavelength dependent information, $\tau_{\alpha
2}' \sim q^{\nu'_2}$; from the fits $\nu_2' \approx 2.2$ for all appropriate
waiting times as summarised in table \ref{tablexpo}. $\tau_{\alpha 2}'$ was found
to depend only very weakly on $q$.
The exponent $\mu_2'=0.7$ was held constant and $f_q$ was used to give the
correct plateau height.

The significance of these exponents is discussed in the next section.

  \begin{figure}[t]\centering\footnotesize
  \includegraphics[width=6cm,angle=-90]{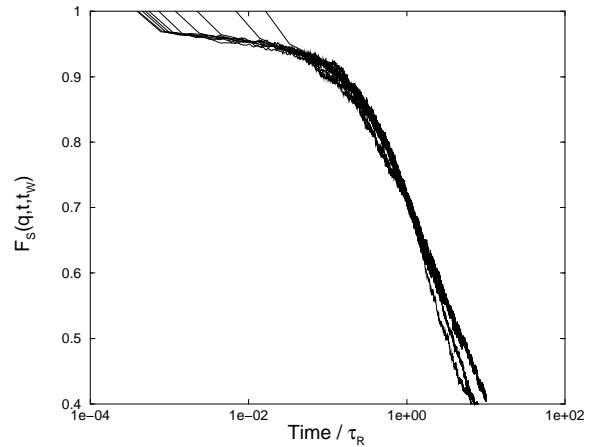}
  \caption{Intermediate scattering curves made to coincide at
  $F_s=0.71$.\label{collapseq625}} 
  \end{figure}

Figure \ref{avintscaaging} shows the aging behaviour of the
gels. The data was all collected at a single $q$ value, $qd=6.25$. The smallest
waiting time considered here is 20 \taur. As the waiting times were increased
from 20 \taur\ up to 2250 \taur\ a lengthening of the $\beta$ plateau was
observed, with the relaxation of the system becoming slower with increasing
waiting time. This is classic aging behaviour for non-equilibrium system and
glasses and has been seen for supercooled LJ systems by Barat and Kob
\cite{barrat97}.

The collapse of the $\alpha$ decay is often observed in experiments on glasses
where one often finds that each set of data can be scaled using a single time
scale which encompasses the whole parameter dependence \cite{gotze}. Hence the
following scaling:

$$ F_s(q,t,t_w) \sim \tilde{F}(t/\tau_{\gamma}) $$ 

with $\tau_{\gamma}=\tau_{\gamma}(t_w)$. Using an arbitrary value of $F_s=0.71$ to
define a relaxation time for each curve $\tau_{\gamma} = t(F_s=0.71)$, the
results were rescaled and reploted on figure \ref{collapseq625} using waiting
times $\ge 100\ \tau_R$. The lower limit was chosen to be at the start of the
last regime seen in figure \ref{totbonds}. As can be seen the data in the range
considered appears to collapse well. This relatively good collapse is
surprising in that the exponent of the stretched
exponential $\mu$ varies with waiting time.

  \begin{figure}[t]\centering\footnotesize
  \includegraphics[width=6cm,angle=-90]{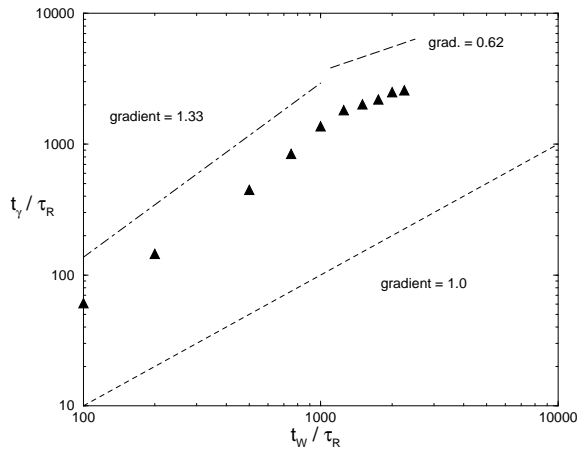}
  \caption{Relaxation times extracted from the rescaling of figure
  \ref{avintscaaging}. \label{agingscaling}} 
  \end{figure}

Figure \ref{agingscaling} shows the dependence of $\tau_{\gamma}$ on $t_w$.
The interpretation of the results in figure
\ref{agingscaling} is slightly ambiguous. It is possible to interpret this data
in two ways. Firstly in terms of a single power law regime $\tau_{\gamma} \sim
t_w^{\xi}$ with $\xi = 1.26$. Secondly
in terms of two power law regimes, the first with $\xi=1.33$ and the second with
$\xi=0.62$, the separation between the regimes being around 1000 \taur. All
of these possible values for $\xi$ are not far from $1$, and similar to
aforementioned results. It was also checked that the same behaviour resulted
if values of $\tau_{\alpha}$ were extracted directly from the fitted stretched
exponentials. The issue of one or two regimes will be discussed
further on.

Another scaling was proposed by M\"{u}ssel and Rieger for the LJ supercooled
fluid results \cite{mussel} and was attempted for comparison. $ F_s(q,t,t_w)
\sim \tilde{F}\{ln[(t+t_w)/\mu]/ln(t_w/\mu)\},$  where $\mu$ is a fit parameter
and plays the role of an affective microscopic length scale. This was proposed
by Fisher and Huse in the context of spin glasses \cite{fisher}. This scaling
was also applied to our data, however it did not produce a successful collapse.

\subsection{Microscopics}

  \begin{figure}[t]\centering\footnotesize
  \includegraphics[width=6cm,angle=-90]{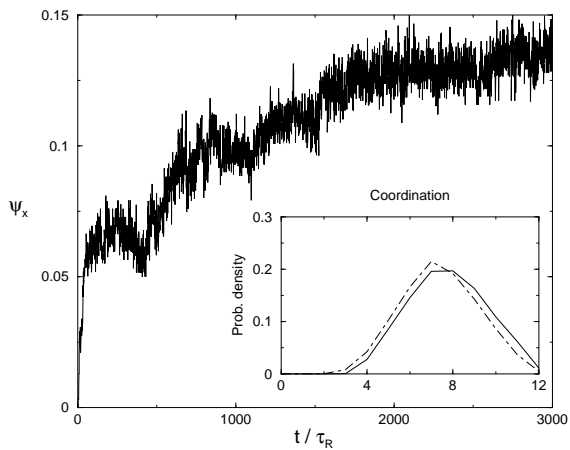} \caption{Crystalline
  Order Parameter. Inset: distribution of particle coordinations in gel; solid
  line: $t_w=3000\ \taur$; dot-dashed line: $t_w=100\ \taur$ \label{xtal}} 
  \end{figure}

  \begin{figure}[t]\centering\footnotesize
  \includegraphics[width=6cm,angle=-90]{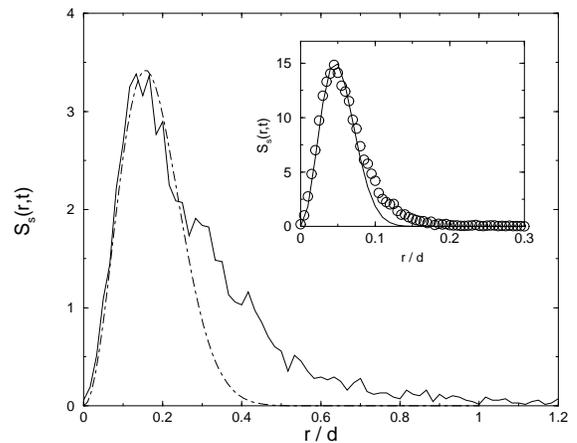}
  \caption{Self part of the Van Hove correlation function. Main graph: long
  time interval, $t_w=3000\ \taur$. Inset: short time interval, $t_w=10\
  \taur$. The peak area is fitted with a gaussian, shown as a dot-dashed line.
  \label{vanhove}} 
  \end{figure}

  \begin{figure}[t]\centering\footnotesize
      \includegraphics[width=6cm,angle=-90]{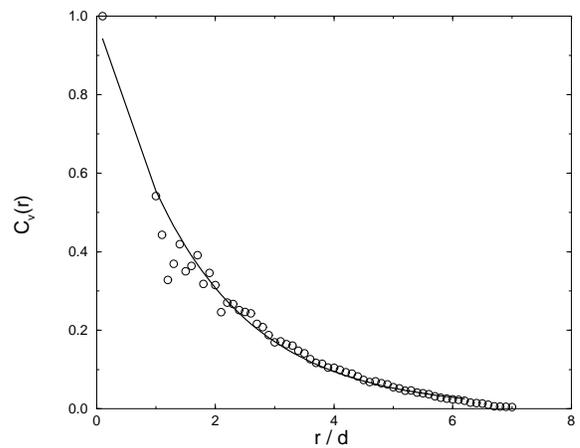}
  \caption{Displacement correlation function, with an exponential fit.\label{vcor}}
  \end{figure}

Over the period of time considered, it is seen that although the average
numbers of bonds in the system increases, there is little increased
crystalinity as shown by the local order parameter, denoted $\psi_x$ (main part
of figure \ref{xtal}). The local crystaline order parameter is defined as the
fraction of of particles which are 12 coordinated or neighbouring a 12
coordinated site. The inset of figure \ref{xtal} shows the probability density
distribution of the coordination of the particles in the system. The initial
crystalinity, which is present very soon after the quench, is just the result
of the high initial colloid volume fraction together with the strong local
order, which is itself due to the lack of angular rigidity of the bonds. There
is therefore no ambiguity about a gel with a non-zero crystalline local order
parameter. At later times an increase of the average position of the peak is
seen but no significant increase in crystalinity. This already rules out one of
the possible hypothesises for the aging of the gel, namely slow growth of
crystals around the crystallites present from formation.

In order to obtain a better understanding of the relaxation mechanisms at work 
the particle dynamics are sampled in real space through the Van Hove
correlation function. S(r,t) gives the probability of finding a
particle at time $t+\delta t$ at a distance r from its position at time t. Due
to the non-ergodicity it is not possible to time average this correlation
function either. However the accuracy of this result was checked by averaging
over a time interval that was short compared by the typical $\alpha$ relaxation
time scale. Figure \ref{vanhove} show a pair of typical results; they were
calculated for $t_w = 3000\ \taur$ and $\delta t=50\ \taur$ and $3000\ \taur$.
For these parameters we expect to be in the $\beta$ plateau for the former time
interval and in the $\alpha$ relaxation regime for the latter. Indeed, the
peaks of the distributions are around $0.05d$ and $0.16d$ which are consistent
with earlier deductions about those regimes. Also shown on figure \ref{vanhove}
are the Gaussian fits for the peaks and the regions around them. The
difference in the behaviour of the dynamics in those two regimes can here be
seen more intuitively. For times up to and including the $\beta$ plateau, the
gel particles' displacements are nearly Gaussian, consistent with a diffusion-like
mechanism. The main part of the plot shows the displacements during the
$\alpha$ relaxation regime, the distribution is very much broader with the
longer displacements not fitting within a Gaussian. The displacements found
after the peak follow a power law $S(r,t) \sim r^{-f}$, with
$f=2.8 \pm 0.4$.

The strongly non-Gaussian displacements rule out a diffusive mechanism, and
hint at cooperative motion. This can be seen visually when colour coding
particle displacements on a 3D representation of the gel (not shown here).
Clusters of particles are observed to be displaced by similar amounts, in fact
this was also true when colour coding displacements which were less then the
peak value. In order to show the presence of collective displacements clearly, a
velocity correlation function was calculated:

\begin{equation}
C_v(r)=\frac{1}{NW}\sum_{ij}^N < \delta(r-(r_i-r_j)) \vec{v}(r_i) \cdot 
\vec{v}(r_j) >,
\end{equation}

Where $W=\sum_{i}^{N} <\vec{v}(r_i) \cdot \vec{v}(r_i) >$ is used to normalise
$C_v$. This is shown in figure \ref{vcor} for $t_w = 4000 \tau_R$ and
velocities calculated over an interval of $\Delta t = 50\ \tau_R$. The data are
fitted with a an exponential allowing a characteristic correlation length $r_v
= 1.9d$ to be extracted. An interesting point was that the results for the
velocity correlation function, for all times beyond the $\beta$ relaxation, were
practically unchanged, thereby showing that whatever the relaxation mechanism,
the same degree of cooperativity is present. 

  \begin{figure}[t]\centering\footnotesize
      \includegraphics[width=6cm,angle=-90]{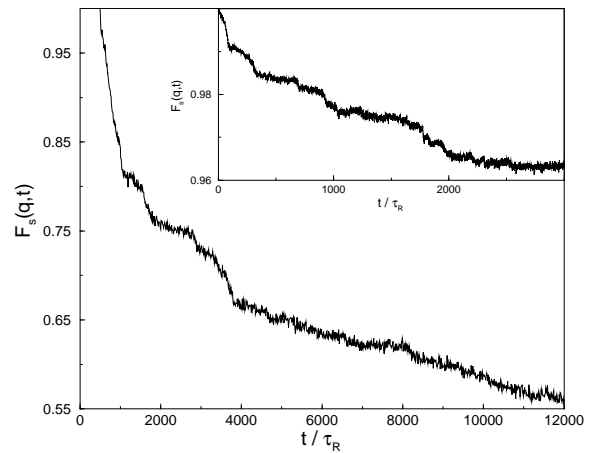}
  \caption{Close up the intermediate scattering function; $qd=6.25$, $t_w=500\ 
  \taur$. Inset: same for a 500 particles system.\label{intscasteps}}
  \end{figure}

  \begin{figure}[t]\centering\footnotesize
  \includegraphics[width=6cm,angle=-90]{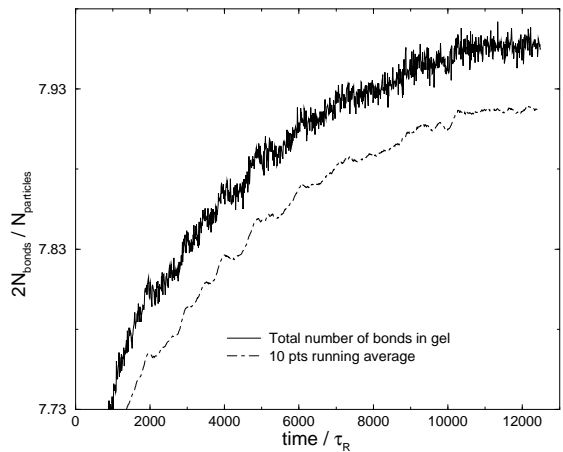}
  \caption{Close up of the scaled total number of bonds in the system and a 10
  point running average.\label{bondsclose}} \end{figure}

Now it is known that particle displacements are spatialy correlated, their
temporal distribution can be studied. For this we calculated the intermediate
scattering function for a single gel, with no averaging procedure. In figure
\ref{intscasteps} the $\beta$ relaxation and the plateau are not apparent due
to the linear time scale which is useful for comparison with other microscopic
results. It is apparent that the relaxation processes at work are not evenly
distributed in time. A succession of plateaus and sharp drops are observed.
Although the displacements are cooperative that this
function is averaged over all particles in the system, therefore a sharp drop
in the averaged function is a sign of major cooperative rearrangement. For a
smaller sized system where a greater fraction of the particles is included in
any given strand the plateaus are closer to the horizontal and the fast
decorrelations are sharper (see inset of figure \ref{intscasteps} for a 500
particle system). 

Following on from the localisation in time and in space of cooperative motion,
we found similar non-uniform behaviour for the total number of bonds in the
system, a close up of which is shown on figure \ref{bondsclose} (scaled to show
the average number of neighbours). A running average has been used to better
show the underlying trends of the data. Here again fast increases are
interspaced with plateaus and short time decreases. Qualitatively we can see
that both the size and gradient of the rapid increases tend to decrease as time
goes on. This was also true of $F_s$ and the same behaviour is apparent in
closeups of the energy (not shown here). It is consistent with a slowing down
of relaxation processes as the gel ages. 

\begin{figure}[t]\centering\footnotesize
      \includegraphics[width=6cm,angle=-90]{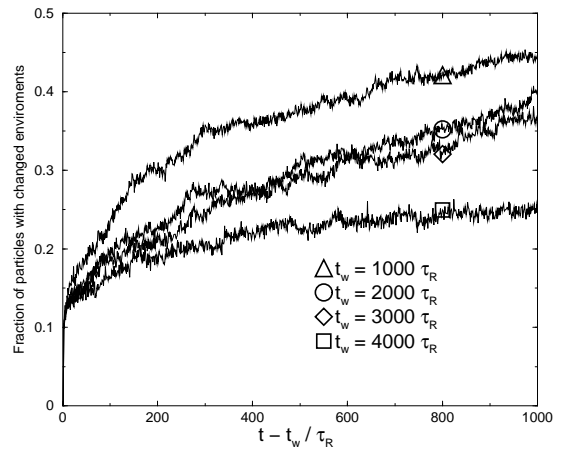}
  \caption{Fraction of particles which change their nearest neighbour
  environment for increasing waiting times.\label{env}}
  \end{figure}

Considering that the particles are strongly localised, it was surprising to see
substantial decreases in the total number of bonds in the system. The lifetime
of a bond based on a simple diffusion over a barrier argument is $\tau_{bond}
\approx \tau_R\ e^{-U/k_BT} \approx 3100\ \tau_R$ for a pair of particles
interacting with this potential. When taking into account the average number of
neighbours within interaction range, one would not expect to find many broken
bonds in any given time interval. Due to the finite range of the potential, it
is possible to define bonds exactly as being within the interaction distance;
therefore, we used a procedure that kept track of all particles that have had
broken or created bonds within a given time interval. Remarkably, this shows
that a high fraction of the particles break and create bonds during the
evolution of the system. Nine percent of the particles were found to have
broken bonds (slightly higher for the created bonds) when comparing
configuration files $500\ \taur$ apart, this is much higher than what would be
expected from figure \ref{bondsclose}. This number is probably an underestimate
since it does not include bonds broken and recreated between the same pair of
particles over that time interval. Over any sizable time interval there is
always a surplus of created bonds. In order to illustrate the effect of the
dynamics we show on figure \ref{env} the proportion of particles that change
their near neighbour environments compared to a reference configuration chosen
every 1000\ \taur. There are three points of importance. There is a short time
interval where about $5-7\%$ of the particles change environment within a
couple of \taur. This regime is independent of the reference configuration and
corresponds to the $\beta$ decay. Its independence suggests that there is a
continuous process of breaking and creating bonds at short times. This will be
further discussed in the next section. Figure \ref{env} also has a second
regime which is waiting time dependent and mostly follows the trends of the
aging behaviour, the rate at which new particles change environment slowing
down with increasing waiting time. The third point to notice is the absolute
values. For the top curve the changes in the gel involve $27\%$ of the
particles, a huge number considering the relatively small distances moved by the
particles.

  \begin{figure}[t]\centering\footnotesize
      \includegraphics[width=6cm,angle=-90]{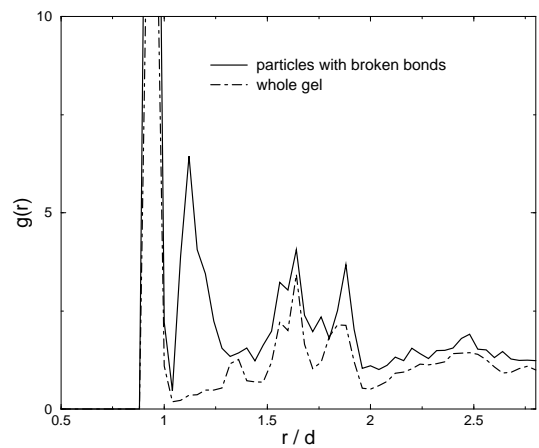}
  \caption{Radial Distribution function done with all the all the particles in
  the gel and by selecting particles with broken bonds.\label{grbroken}}
  \end{figure}
  
The presence of particle displacements that are greater than the contribution
from the elastic modes, of correlated dynamics and a high fraction of particles
changing environments, meant we expected to find some correlations in the
position of those particles which have broken bonds or changed environment.
However breaking bonds in a concentrated system is always going to lead to some
correlation between the particles; to see this imagine pulling away one of the
particles of a tetrahedron, the three particles left behind have all had a
broken bond but are highly correlated in space.
A visual inspection of the gel with highlighted particles where bonds had been
broken confirmed the suspected correlation; we also calculated a pair
correlation function $g_b(r)$ by selecting the particles that had broken bonds
in a given time interval (figure \ref{grbroken}), and comparing it to the
normal pair correlation $g(r)$. The correlation shows up in $g_b(r)$ as a
higher first neighbour peak and a large peak beyond it which is found at the
limit of the interaction potential, which is no surprise since it is by
definition the distance at which one considers a bond to be broken. In terms of
the previously given argument about pulling apart a tetrahedron, the higher
first peak is due to the strong correlation of the particles left behind, while
the second peak is due to the particles that are just beyond the interaction
distance, \ie\ the ones that have just broken off. This second peak at the
limit of the potential is probably related to the initialy rapid increase seen
in figure \ref{env}, since it most likely shows that a lot of particles are
hovering near the edge of the potential well, accounting for much of the bond
breaking as they pop in and out when the strand is deformed by thermal motion.

Exactly the same observations were made about the particles that had created
bonds in any reasonable time interval.

Since it was the case in other studies \cite{cipe1}, one possible driving force
for the aging of the gel could be the internal stresses. It is known that gels
formed in Periodic Boundary Condition have non-zero normal stresses
\cite{heyes98}, and experimentaly a gel which has been detached from the walls
of its container shrinks macroscopicly \cite{cipe1}; figure \ref{strav} shows
the average of the normal stresses, raw data in black and a running average in
white. The raw data show the high frequency noise due to the brownian
fluctuations, but the presence of underlying low frequency fluctuations is
revealed by the running average. The general trend is that of a decrease, and
there are large fluctuations in the average, although not as large as any of
the fluctuations in the raw data. A spatial stress correlation function,
$C_{\sigma}(R) = <\sigma(r)\sigma(r+R)>$ was calculated but was found to show
only very low spatial correlations.

The distribution of bond lengths in the gel was found to be quasi-equilibrium
with only a small discrepancy at distances beyond the potential minima due to
the normal stresses present in the system, as can be seen on figure \ref{gibbs}. 

\section{Discussion and Conclusions}

This study of the dynamics of gels formed from colloids interacting via a
depletion potential has showed that complex behaviour arises during the aging.

  \begin{figure}[t]\centering\footnotesize
      \includegraphics[width=6cm,angle=-90]{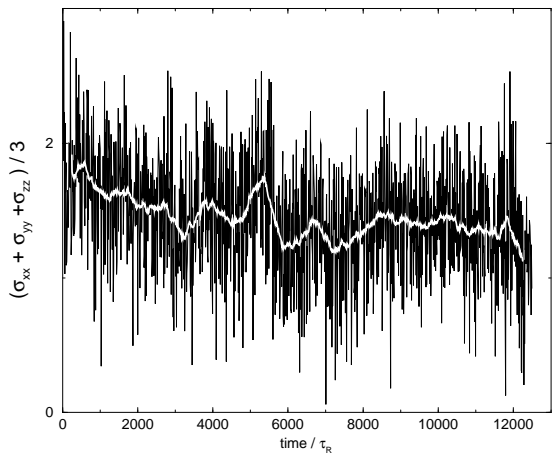}
  \caption{Averaged normal stress tensor components (negative of pressure); raw
  data in black, 20 point running average in white.\label{strav}}
  \end{figure}

  \begin{figure}[t]\centering\footnotesize
      \includegraphics[width=6cm,angle=-90]{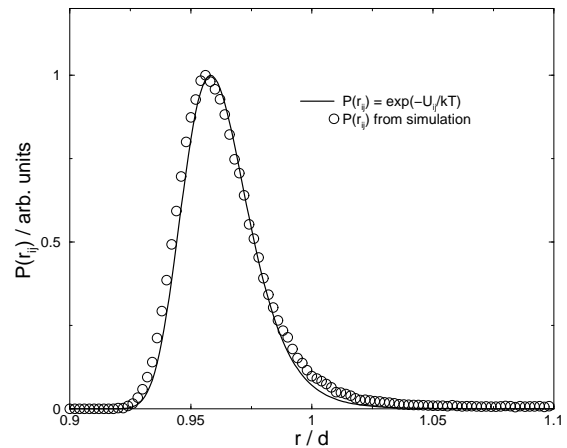}
  \caption{Distribution of bond lengths in gel (circles). Black line is
  $P = \exp(-U_{ij}/kT)$. Recall that range of the potential is $1.1d$.
  \label{gibbs}}
  \end{figure}
  
It was found that the total potential energy of the system could be described
in terms of a power law with an exponent $b=0.3$ which is an intermediate value
to what was obtained by Barrat and Kob for a Leonard Jones glass (b=0.14)
\cite{barrat97} and by Parisi \cite{parisi97} (b=0.7) for a MC generated glass.
However, the total number of bonds in the system was found to grow as three
separate logarithmic regimes. Some of this three regime structure was also
present in the energy plot demonstrating the difficulty in describing
accurately these very slow growths/decays. The impossibility in calculating
these quantities in experiments make verification difficult if not impossible.
However, whether interpreting the slow growth of the number of bonds as a power
or logarithmic law, the presence of three distinct regimes is of much interest
and demonstrates the complex behaviour associated with the formation and
consolidation of the gel. We know from previous work that the time to percolate
is less than $1\ \tau_R$ \cite{sogaperc}. Following this one would expect some
fast consolidation of the network during which thin strands coarsen.  
Since diffusion processes occur, by definition, on the time scale
of $1\ \tau_R$, one would expect complex and fast rearranging at relatively
short times and short length scales as was seen by Haw \etal for 2D colloid
simulations where compactification was obvious at short times \cite{haw95}.
However, given that we know that the end product of this is not at
thermodynamic equilibrium, these fast rearrangement processes  must be
inactive at intermediate to long times (otherwise the equilibrium phases would
result). Although all the regimes seen on figure \ref{energy} occur after 
percolation, it is beyond the scope of this paper to explain all three, our
attention is here focused on the last one for several reasons: it is expected
that the behaviour maybe less complex in this region; these time scales are
more accessible to experiments thereby providing interesting comparisons on the
simulation results and insights that may be tested on real systems. In order to
study the aging and not the formation/consolidation behaviour the work showed
here focuses on the behaviour in the regime from $100\ \tau_R$ after quenching
onwards. 

As an aside it is worth mentioning that the noise in the energy of the gel goes
as ${FT(U)}^2 \approx 1/f^{1.4}$, this value being closer to the 1/f flicker
noise than the more usual $1/f^2$ white noise. This has been linked in the
literature to Self Organised Criticality and to cooperative behaviour in
complex systems. However this needs to be studied further before conclusions
can be drawn from it.  

The self part of the ISF calculated for this system displayed the classic two
step relaxation. Both of these relaxations were found to be well approximated
by stretched exponentials with a plateau between them. 

The age invariant short time dynamics, corresponding to the beta relaxation was
interpreted in terms of the Krall Weitz model for the decorrelation due to the
elastic modes present in a gel.  Although they first used it for describing low
volume fraction fractal gels, the physical reality of the model is still
present even though the gel is not a fractal object. This model has been shown
to work well in describing gels at higher \vfc\ by Romer \etal \cite{romer00}
with reduced values of $\delta^2$ and $\tau_c$ reflecting on the compactness of
the gels.   

The mean square displacements and the beta relaxation were fitted using
equations \ref{deltar} and \ref{stretched}; the value of the exponent was
determined to be $p=0.4 \pm 0.1$ which is distinctly smaller than Krall and
Weitz ($p=0.7$) for their low \vfc\ polystyrene gel \cite{krall98}, Romer \etal
also found $p=0.7$ for a high \vfc\ polystyrene gel \cite{romer00}. However
Solomon and Varadan found $p=0.5$ for a low \vfc\ thermoreversable gel of
silica particles \cite{solomon01}. In the Krall Weitz model $p$ is related to
the length scale dependent spring constant for fractal clusters via $\kappa(s)
\sim s^{-\beta}$ and $p = \frac{\beta}{\beta+1}$. From computer simulation
$\beta \sim 3.1$ for DLCA \cite{meakin84}; from Krall and Weitz, and Romer
\etal $p=0.7$, $\beta=2.3$; whereas from Solomon and Varadan $p=0.5$, therefore
$\beta=1.0$ for thermoreversable gels which the authors believed had some
angular rigidity on the local level due to surface roughness and a different
aggregation mechanism leading to a locally denser gel. Furthermore, previous
studies by one of the authors on 2D gel networks where trimers provided some
angular rigidity \cite{west94}, showed that for such systems $\beta \sim
1.1-1.3$. All of the previous suggests, as proposed by Solomon and Varadan, the
the elasticity exponent $\beta$ is strongly dependent on the angular rigidity
of the local network. Here $p=0.4$ therefore $\beta = 0.66$ suggesting a
structure that only slowly becomes more compliant with increasing length scale.
There are several factors influencing this in these simulations. Firstly there
is no imposed angular rigidity for the colloid interaction (which is different
from experimental particles which may be rough or deform under the strength of
the interactions). Secondly the colloid volume fraction used here is \vfc=0.3,
much higher than most experiments; and thirdly, the interaction range is very
small ($0.1d$), this has been shown by Heyes and co-workers \cite{heyes98} to
affect the structures obtained when forming the gels. Short interaction
distances resulted in the creation of gels where the local structure was very
compact, for example it can be seen as a split second peak in g(r), whereas
when longer ranged potentials were used `fluffier' structures resulted. The
former is the case in this study, and it is obvious that a thicker strand of
colloids has a higher angular rigidity. Furthermore, this third point is
strengthened by the high \vfc, and probably negates the first point since the
lack of angular rigidity of a single bond becomes irrelevant if there are no
singly connected particles in the strand. This  explains the high angular
rigidity of this structure and hence the low values of $p$ and $\beta$. An
interesting point that arises out of this discussion is that for a given
maximum depth of potential, the rigidity of the gel may be influenced strongly
by the range of the potential. This may explain why for the higher volume
fraction gels of Romer \etal $p=0.7$, the same as for the Krall and Weitz's low
volume fraction gels, although what the range of the
interactions present in the aforementioned experimental systems was not clear.

This physical picture is confirmed by direct visualisation of the gel which
shows the clusters of particles moving relative to each other on time scales of
order $1\ \taur$. Furthermore this is consistent with the high number of
particles which change environment during such time intervals. This is a direct
consequence of clusters in relative motion within a relatively stiff structure,
thereby causing many bonds to break and reform.

The length scale dependent plateau height as characterised by $f^s_q$ was found
to have a very large $q$-width. This result is very similar to Puertas \etal
\cite{puertas02}, who simulated the approach to gels in concentrated fluids
near the glass transition. They obtained a similar broad dependence showing
again the strong confinement of the particles due to the short range
potentials. This extends their results to a lower \vfc\ than they have used.
A further distinction is that this result is at a point on the phase diagram
deep in the gel state whereas they were in the fluid leading up to the gel
state. 

The $\alpha$ relaxation for waiting times $\ge 500\ \taur$ was found to be well
described by a stretched exponential $F_s(q,t,t_w)= A\
exp(-((t-t_w)/\tau_{\alpha})^\mu)$ where the exponent was always  $\mu < 1.0$.
This type of law is often observed and results from the presence of relaxation
modes with a wide distribution of time scales. This can be written as a Laplace
transform over the distribution of relaxation modes \cite{gotze}: $F(t) =
\int^{\infty}_{0}\rho(\gamma)e^{-\gamma t} d\gamma$. The exponent of the
stretched exponential is strongly dependent on the shape of the distribution
$\rho(\gamma)$, wider distributions leading to lower exponents. In this study, 
the variation of the exponent of the stretched exponential, $\mu = 0.45
\rightarrow 0.72$ as $t_w = 500 \rightarrow 2250 \taur$ clearly indicates a
reduction of the width of the modes distribution. Furthermore the observation
that the variation in $\mu$ is fast for small waiting times and levels to $\mu
\sim 0.7$ for longer waiting times and also the observation of the change in
behaviour for $t_w < 500$ where the $\alpha$ relaxation appeared to have a
`fast' decay followed by a `slow' one, making it impossible to fit a stretched
exponential, all point to the possibility of $\rho(\gamma)$ being a bimodal
distribution where the short time modes are widely separated from the long time
modes. Hence the success of using a function containing two stretched
exponentials to fit the decay of the alpha relaxation up to waiting times of
$t_w=100\ \taur$. In this scheme each of the exponentials describes one part of
the distribution. This behaviour was reproduced by numerical integration of the
Laplace transform for distributions of modes made of two widely separated
Gaussians. All of the correlation function data can then be explained. $\mu$
and $\mu_2'$ describe the distribution of the `slow' modes,
$\mu=\mu_2'=0.7\pm0.2$ for all waiting times larger than the time for
the slowest of the `fast' modes to relax. $\mu_1'$ which describes the
distribution of `fast' modes increases towards 1.0 as the distribution of the
modes that have not yet relaxed narrows. For intermediate to long waiting times,
a single stretched exponential is recovered.

The $q$ dependence of the characteristic relaxation times $\tau_{\alpha},
\tau_{\alpha 1}', \tau_{\alpha 2}'$, extracted from the fits also contain
information about the processes at work. $\nu$ at long waiting times, and
$\nu_2'=2.2\pm0.1$; for the latter exponent this was true for all the
waiting times back to $t_w=100\ \taur$. This is very close to a diffusive law,
$\tau \sim q^{-2}$, yet with a consistent deviation. $\nu_1'$ was found to vary
only very weakly with $q$ highlighting the fact that different mechanisms are
responsible for the 'fast' and 'slow' relaxations. It also shows that these
'fast' modes are not related to the Krall Weitz elastic modes which showed up in
experiments as $\tau \sim q^{-2}$.

Comparing these relaxation processes with the much more studied glasses, the
$\alpha$ relaxation behaviour is more complex in this case as opposed to
homogeneous supercooled liquids \cite{barrat97}. It may not be surprising that
there should be some processes that would be specific to gels, considering their
spatial inhomogeneity.

This framework also explains why in experiments a constant value of the
stretched exponential exponent is usually found, since the `fast' modes 
observed here would be inaccessible.

When considering the ISF in non-equilibrium systems, aging can be observed to
occur as a lengthening of the plateau at intermediate times. The response
becomes strongly dependent on waiting time after the quench. This was seen
clearly in our data. By making the curves coincide at one point (or
equivalently using the extracted parameters from the fits), a collapse within a
fairly short time window was observed. It is obvious that since the exponents
of the stretched exponentials are slightly waiting time dependent, the collapse
cannot be perfect over a long time scale, however it is an important feature
displayed by many gels and glasses \cite{puertas02}. The dependence of
$\tau_{\gamma}$ on the the waiting time is explained from the previous
deductions about our system, taking into account the fact that $\tau_{\gamma}$
is nearly equivalent to $\tau_{\alpha}$ for a single stretched exponential
describing the whole decay. The presence of slow and fast modes, and therefore a
change of behaviour once the short modes are no longer present, was observed
again in what appeared to be two different aging regimes in figure
\ref{agingscaling}. It is in fact a confirmation of that change in behaviour
due to the distribution of modes going from bimodal to single peaked. Therefore
the correct value of $\xi$ to compare to experiments is $\xi=0.66$. Previously
$\tau \sim t_w^{\xi}$ with $\xi=0.88$ \cite{barrat97} and $\xi \sim 1.1$
\cite{fisher} have been found for glasses, close to $\xi=1$ which would be
the simple aging case.

The experimental studies of Cipelletti \etal have reported $\nu = 1$, \ie\ 
$\tau_{\alpha} \sim q^{-1}$, meaning the length scale displaced scaling
linearly with time. This was done for both attractive and repulsive systems,
where it was thought in both cases that the relaxation of internal stresses
dominate the dynamics. The theoretical argument from the the same studies also
explained their value of the exponent $\mu=1.5$. The aging behaviour found was
$\tau_{\alpha} \sim t_w^{\xi}$ with $\xi=0.9$ after a period of exponential
growth for the gels \cite{cipe1} and $\xi=0.78$ for the repulsive system
\cite{cipe2}. The very different values of $\mu$ and $\nu$ obtained here point
to major difference the mechanisms of the aging even if the scaling of the
relaxation times may look similar. 

The Van Hove correlation function gives us a more intuitive idea
of the particle displacements in real space. For large observation times and
long waiting times, as is the case in figure \ref{vanhove}, the system is in a
regime where the displacements are greater than the $\beta$ plateau regime.
For comparison with a system with diffusion-like displacements a Gaussian was
fitted to the peak. The very large tail present shows just how non-Gaussian the
dynamics are within the $\alpha$ relaxation regime. The displacements around
the peak and up to $0.25d$ appear to be Gaussian distributed, leaving the
aforementioned large tail. Not all displacements within this tail are due to
the same physical processes. We can distinguish between two types of behaviour.
Firstly single particle movements: particles that break away from the network,
becoming part of the gas phase before rejoining the network somewhere else or
moving along the gel by surface diffusion. These particles make up all of the
displacements greater than $1d$, but the numbers concerned are small (due to the
high coordination of the particles in the gel), \ie\ they could not account on
their own for the $\alpha$ relaxation. Secondly, there are particles which have
moved distances of around $0.25d < r < 1d$, but which never leave the network,
and for which the displacements are correlated in time and space.
We also note that the part of the displacements found after the peak follows a
power law with an exponent $f=2.8 \pm 0.4$. This is, within errors, what was
deduced by Cipelletti \etal and predicted by Bouchaud and Pitard
\cite{bouchaud01}. However, as we have just seen, their system is quite
different being strongly attractive, and other parameters such as the wavevector
dependence are different from this work. Further work is necessary before
conclusions can be drawn from this.

The microscopic data presented also indicates that the aging does not occur
as continuous process but rather as a succession of 'events' which occur on a
time scale which increases with increasing waiting time, as was observed on a
close up of the number of bonds per particle. Indeed it seemed that for some
short time intervals the number of bonds in the system decreased. These events
involve a relatively large number of particles moving cooperatively since they
can even be seen on a correlation function averaged over 4000 particles, and
dominate the gel dynamics for time scales greater than the duration of the beta
plateau.

While the ideas of cooperative dynamics in non-equilibrium systems is not new,
string-like motion of chains of particles moving cooperatively having been
observed in supercooled liquids \cite{donati98}, the behaviour seen here is
more reminiscent of the work of Sanyal and Sood \cite{sanyal98} who studied a
dense binary colloid mixture in which step like displacements were observed.
This was accompanied by some hop back motion as well. It is likely that a site
hopping scenario is responsible for this, whereas back and forth displacements
were not observed in the gel case, although such motion may be present at
shorter times.

Unsurprisingly, these correlated dynamics were linked to a correlation in the
localisation of the particles that have broken or created bonds for a given
time interval. The relatively high number of bonds broken (and created) in the
gel when comparing configurations separated by a fixed time interval is an
indication of the \emph{ease} with which thermal fluctuation of the strands
disrupt the network, causing a high fraction of particles to change environment.
This may be a result of the spatially anisotropic nature of the gel, whereas
large thermal flutuation are more unlikely to occur in an isotropic glassy
system.

Although the low frequency trend in the stress data from figure \ref{strav} is
the result of the microscopic aging `events' that occur in the gel it is now
thought by the authors to be unlikely to be the driving force. While it was a
surprise to find that the stresses were not spatially correlated, this may point
to the fact that forces that can be sustained by the inter-colloidal bonds are
not very high in comparison to the stresses occasioned in the strands by the
thermal fluctuations, even though they suffice to confine the individual
colloid particles to the gel. Under the influence of thermal fluctuations
strands can break bonds relatively easily; this is the only explanation for
the the large number of bonds broken in the system. A further clue which
supports this argument is the relative size of the high frequency stress
fluctuations compared to both the average value of the stress, and the size of
the low frequency fluctuations. In both cases the stresses caused by the
thermal fluctuations are of much higher magnitude. This supports the idea that
the internal stresses cannot be responsible for the aging, and that the low
frequency fluctuations and the general decreasing trend of the average stress
are consequences of the aging, not driving factors.

It is not yet clear why the relaxation times increase with waiting time,
however the mechanisms invoked here involve breaking the network under the
thermal fluctuations. Presumably, the network will break preferentially at the
point at which it is weakest. These weak points between the strands that make
up the network would tend to disappear in time (they will keep breaking until
they do). As the network become devoid of such weak points, we can expect the
dynamics to slow down, thereby causing the aging behaviour. Further work is
underway to clarify this issue.

Comparison of this work with that of Barrat and Kob, reveals great similarities
in the overall properties of the relaxation dynamics in these gels and glasses,
the same aging was seen in the correlation functions, as well as superposition
of the $\alpha$ relaxation regime. However it is clear from this work that the
microscopic  mechanisms by which the aging occurs in the gels are different to
the ones observed in  homogeneous glasses, the dynamics in the latter involving
cooperative motion of strings of particles \cite{donati98}. Similarly,
comparison with Puertas \etal \cite{puertas02}, shows the same overall
behaviour even though their simulations focused on the fluid states leading up
to the gel transition. The relaxation mechanisms (although they did not report
any details) were very probably similar to what was observed in
\cite{donati98}.

The aging behaviour predictions given by non-equilibrium MCT are power law
regimes either side of the $\beta$ plateau, a waiting dependent $\alpha$
relaxation time and violation of the fluctuation dissipation theorem. Although
the predictions are not specifically tested here, our data, at least
qualitatively is not in disagreement with them. We found a stretched
exponential fit to be better than a power law  for the $\alpha$ relaxation but
within a certain region near the plateau, a power law is certainly possible.

The Potential Energy Surface approach lends itself well to the interpretation
of our data, and comparison with aging glasses. In this picture, the gel would
be in a regime in configuration space where the system samples basins of IS
with ever decreasing energy, but with the barriers between Inherent Structures
becoming greater than the thermal energy in the system. The slow hopping over
barriers manifests itself in the gel as network rearrangements, and hence the
behaviour of quantities such as the total number of bonds in the system. 

Based on the work of Kob \etal
\cite{kob00}, who have showed that an effective time dependent temperature
can be defined for an aging system, Sciortino and Tartaglia \cite{sciortino01}
have proposed an out of equilibrium thermodynamics framework based on the IS
description of the PES. In this picture, the free energy is separated into a
basin term and a configuration term reflecting the separation of time scales in
supercooled fluids. In the case of a temperature jump the system equilibrates
to the bath temperature in the same basin rapidly. The exploration of
configuration space takes place on a much longer time scale, hence the aging
observed as the system lowers its configurational energy. This description is
fitting for our system where figure \ref{gibbs} shows that the system is in
quasi thermal equilibrium while of course not being in thermodynamic
equilibrium. For the correlation functions calculated for a Leonard Jones
Binary Mixture, Sciortino and Tartaglia observed stretched exponentials for the
equilibrium fluid, but logarithmic decays for the out of equilibrium LJ fluid.
Note that this is in fact an MCT prediction for quenches near an $A_3$ point.
Once again superposition of the $\alpha$ relaxation held and relaxation
time scales increasing with $t_w$ were found. 

The different regimes seen the total number of bonds in the gel (figure
\ref{energy}), can then be interpreted in terms of the PES, much in the way of
\cite{sciortino01}and \cite{kob00}. After the
instantaneous quench the system is still fluid as aggregation begins (0 to 10\ 
\taur), this is followed by ridge dominated dynamics (10 to 100\ \taur) and
finally the aging behaviour of the last regime dominated by the slow
sampling of IS of lower and lower energy.

The similarities of this work on supercooled fluids with ours is striking and
the only difficult point concerns the shape of the $\alpha$ relaxations.
Whereas studies of the PES of liquids found logarithmic decays, we had two
regimes of stretched exponentials. It is possible that is it more strongly
influenced by the microscopic mechanisms than what was previously thought.
Those mechanisms could well be different due to the localisation of the
particles within stands rather than being distributed nearly isotropicaly. The
thermal fluctuations in strands being enough to break the network thereby
causing the aging even though thermal fluctuations of a single particle are very
rarely enough for them to break away from a strand.

This highlights a problem concerning general theories of the glass transition,
if the microscopic details that bring about the final relaxation effect the
macroscopic observations, as was seen when comparing our results with the
stress driven systems of Cipelletti \etal \cite{cipe1} and Solomon \etal
\cite{solomon01} and the aforementioned results for supercooled fluids, then a
theory like MCT which works with averaged quantities is unlikely to distinguish
between them. On the other hand, there seems to be a greater chance to
incorporate different microscopic relaxation theories when using the PES
approach in order to estimate the time to explore different basins, and
understand its relation with the aging time. 

\section{Conclusions}
To finally conclude this study, we can say that simulations now appear to be a
reasonable way to study the short term aging of soft materials. It has allowed
us to study the intermediate scattering function of an aging gel in order to
compare it to experimental studies while at the same time to understand the
microscopic mechanisms responsible for the aging. The conclusion is that for
this weakly attractive, gel-forming, system the aging is caused by discontinuous
network rearrangements due to the thermal fluctuations. It was also possible to
investigate and further strengthen the relationship between gels and glasses.
A possible extension of this study would concern the possibility of reproducing
experimental results for a more stress driven system to check the relationship
between the $\alpha$ relaxation and the microscopic mechanisms.

\section{Acknowledgements}

RA acknowledges the support of EPSRC and Colworth labs (Unilever Research)
for funding; he would also like thank H. Lekkerkerker, W. Poon, A. Puertas, M.
Fuchs, L. Starrs, L. Cipelletti, and members of the Polymers and Colloids group
for many useful discussions.

\bibliography{paper}

\end{document}